\documentclass[aps,prl,twocolumn,showpacs,superscriptaddress,nofootinbib,groupedaddress]{revtex4-1} 
\usepackage{graphicx}
\usepackage{epsfig}
\usepackage{tabularx}
\usepackage{booktabs}
\usepackage{amssymb}
\usepackage{amsmath}
\RequirePackage[colorlinks,citecolor=blue,urlcolor=blue,linkcolor=blue]{hyperref}
\usepackage{dcolumn}
\begin{document} 
\preprint{PRD}

\title{Generating nonzero $\theta_{13}$ without breaking the $2$-$3$ symmetry of neutrino mass matrix }

\author{S.~Roy}
 \email{meetsubhankar@gmail.com}
\affiliation{Department of Physics, Gauhati University, Guwahati, Assam -781014, India
}%
\author{N.~N.~Singh}
 \email{nimai03@yahoo.com}
\affiliation{Department of Physics, Manipur University, Imphal, Manipur-795003, India
}%
\date{\today}

\date{\today}


 
\begin{abstract}
The prediction of vanishing reactor angle was thought to be a signature of $2$-$3$ symmetry of neutrino mass matrix. But the present study addresses certain interesting facts related with $2$-$3$ symmetry which are not addressed so far. The investigation highlights that $\theta_{13}=0$, corresponds to a very special case in association with $2$-$3$ symmetry and to engender a non-zero $\theta_{13}$, the breakdown of $2$-$3$ symmetry is not essential.
\end{abstract}

\maketitle

\section{Introduction}
In the last few years, when the reactor angle ($\theta_{13}$) and the atmospheric mixing were expected to be zero\,\cite{Apollonio:1997xe} and maximal respectively, the $2$-$3$ symmetry was anticipated to be a perfect symmetry for neutrino mass matrix \cite{Fukuyama:1997ky,Ma:2001mr,Lam:2001fb,Balaji:2001ex}. Several models based on this symmetry were framed. But, as the recent data proclaims a nonzero finite $\theta_{13}$\,\cite{Capozzi:2016rtj,Forero:2014bxa,Gonzalez-Garcia:2015qrr}, the relevance of $2$-$3$ symmetry turns weaker.  Here we emphasize on the fact that believing vanishing $\theta_{13}$ or $|\nu_{3}\rangle=(0, 1/\sqrt{2},1/\sqrt{2})^T $ as unique signature of $2$-$3$ symmetry is certainly misleading. In the present article we revisit and generalize the finer facets of the $2$-$3$ symmetry unattended so far, from a simple model independent stand. The article attempts to show that a non-zero $\theta_{13}$ is not at all foreign to an exact $2$-$3$ symmetric framework and to engender the former, the break down of the latter is not at all mandatory.   

The neutrino mass matrix ($m_{\nu}$) in general shelters nine physical parameters which are the three neutrino mass eigenvalues ($m_{i=1,2,3}$), three mixing angles ($\theta_{12}$, $\theta_{23}$, and $\theta_{13}$) and three complex phases: one Dirac type ($\delta$) and two Majorana phases ($\alpha$ and $\beta$). The Pontecorvo-Maki-Nakagawa-Sakata (PMNS) matrix ($U$)\,\cite{Maki:1962mu} contains the information of all the three mixing angles and three complex phases. Following standard parametrization, one can express,
\begin{eqnarray}
&& U = \nonumber\\
&& \begin{small}
\left(
\begin{array}{lll}
 c_{12} c_{13} \pmb{\text{}} & c_{13} s_{12} \pmb{\text{}} & e^{-i \delta } s_{13} \pmb{\text{}} \\
 -c_{23} s_{12} \pmb{\text{}}-e^{i \delta } c_{12} s_{13} s_{23} \pmb{\text{}} & c_{12} c_{23} \pmb{\text{}}-e^{i \delta } s_{12} s_{13} s_{23} \pmb{\text{}} & c_{13} s_{23} \pmb{\text{}} \\
 s_{12} s_{23} \pmb{\text{}}-e^{i \delta } c_{12} c_{23} s_{13} \pmb{\text{}} & -e^{i \delta } c_{23} s_{12} s_{13} \pmb{\text{}}-c_{12} s_{23} \pmb{\text{}} & c_{13} c_{23} \pmb{\text{}} \\
\end{array}
\right)
\end{small}\nonumber\\
&& \quad \quad\times\quad P
\end{eqnarray} 
where, $P=diag \lbrace e^{i\alpha},e^{i\beta},1\rbrace$\,\cite{Agashe:2014kda}. In the above expression of $U$, the $s_{ij}$, $c_{ij}$s represent $\sin\theta_{ij}$ and $\cos\theta_{ij}$ respectively. The matrix $U$ in fact diagonalizes, $h=m_{\nu}m_{\nu}^{\dagger}$. 
\begin{equation}
\label{h}
diag\lbrace m_{1}^2,m_{2}^2,m_{3}^2\rbrace =U^{\dagger}.h.U.
\end{equation} 
The $h$ does not contain the information of the two Majorana phases. 

In general, $U$ suffers contribution of both charged lepton and neutrino sectors. One can express, $U=U_{lL}^{\dagger}.U_{\nu}$. Here, $U_{lL}$ is the unitary matrix that diagonalizes, $h'=M_{l}.M_{l}^{\dagger}$, where $M_{l}$ is the charged lepton mass matrix. There are many Grand unified theory based models\,\cite{Pati:1974yy,Elias:1975kf,Elias:1977bv, Blazek:2003wz,Dent:2007eu} that suggests the texture of $M_{l}$ to be of down quark mass matrix type. This allows one to choose $U_{lL}$ as exactly or near to $V_{CKM}$ type\,(Where, $V_{CKM}$ is the Cabibbo-Kobayashi-Maskawa matrix)\,\cite{Wolfenstein:1983yz}. It is a well known fact that $V_{CKM}$ is almost an identity matrix. Hence as a \textit{first approximation}, one can choose the basis such that $U_{lL}\approx I$, or in other words a basis where $M_{l}M_{l}^{\dagger}$ is diagonal. Hence, under this approximation the PMNS matrix can be directly identified with the neutrino mixing matrix. Throughout our discussion we shall stick to this set up.

If the neutrinos are Dirac type, the existence of two CP violating phases $\alpha$ and $\beta$ are irrelevant. On the other hand, the Majorana nature demands the significance of all the three phases. In this article, we assume the neutrinos as Majorana particle and hence the general neutrino mass matrix\,($m_{\nu}$) takes a symmetric texture, 
\begin{equation}
m_{\nu}=\begin{pmatrix}
A & B & E\\
B & C & D\\
E & D & F
\end{pmatrix},
\end{equation}
and can be diagonalized in the following way,
\begin{equation}
\label{mnu}
diag\lbrace m_{1},m_{2},m_{3}\rbrace =U^{\dagger}.m_{\nu}.U^{*}.
\end{equation}

The Bi-maximal\,(BM)\cite{Mohapatra:1998ka} or the popular Tri-Bimaximal\,(TB)\,\cite{Harrison:2002er,Xing:2002sw} mixing are certain mixing patterns which  are relevant from group theoretical point of view, and they define $U$ (upto the Majorana phases) in the following way,
\begin{equation}
\label{U1}
U=\left(
\begin{array}{ccc}
 c_{12} & -s_{12}  & 0 \\
 \frac{s_{12} }{\sqrt{2}} & \frac{c_{12}}{\sqrt{2}} & -\frac{1}{\sqrt{2}} \\
 \frac{s_{12}}{\sqrt{2}} & \frac{c_{12}}{\sqrt{2}} & \frac{1}{\sqrt{2}} \\
\end{array}
\right),
\end{equation}
with,
\begin{eqnarray}
s_{12}=\sqrt{\frac{1}{2}},\quad c_{12}=\sqrt{\frac{1}{2}}\quad(BM)\nonumber\\
s_{12}=\sqrt{\frac{1}{3}},\quad c_{12}=\sqrt{\frac{2}{3}}\quad(TB).\nonumber
\end{eqnarray} 
The common mixing angles are: $\theta_{23}=45^{0}$, $\theta_{13}=0$. The mixing angle, $\theta_{12}$ is predicted to  be $45^{0}$ and $35.26^{0}$ by BM and TB mixing respectively. Both of them define $|\nu_{3}\rangle$ as, $|\nu_{3}\rangle=(0, 1/\sqrt{2},1/\sqrt{2})^T $ and this characteristics associates the neutrino mass matrices from both the mixing patterns to assume a $2$-$3$ symmetric texture as such,
\begin{equation}
m_{\nu}=\begin{pmatrix}
A & B & B\\
B & C & D\\
B & D & C
\end{pmatrix}.
\end{equation}
Again, a different parametrization of $U$ as such,
\begin{equation}
\label{U2}
U=\left(
\begin{array}{ccc}
 c_{12} & s_{12}  & 0 \\
 -\frac{s_{12} }{\sqrt{2}} & \frac{c_{12}}{\sqrt{2}} & \frac{1}{\sqrt{2}} \\
 \frac{s_{12}}{\sqrt{2}} & -\frac{c_{12}}{\sqrt{2}} & \frac{1}{\sqrt{2}} \\
\end{array}
\right),
\end{equation}
leads to a texture with additional minus sign before $(m_{\nu})_{13}$,
\begin{equation}
m_{\nu}=\begin{pmatrix}
A & B & -B\\
B & C & D\\
-B & D & C
\end{pmatrix}.
\end{equation}
The difference in the two textures is attributed to the the choice of $U$ in Eqs\,(\ref{U1}) and (\ref{U2}). In the first case, the rotations are assumed to be clockwise, while the latter assumes the same as anti-clockwise. The properties of $2$-$3$ symmetry with which we are familiar so far are summarized below.
\begin{itemize}
\item[(a)]It says, $\theta_{13}=0$,\,$\theta_{23}=45^0$,\,$\theta_{12}\rightarrow$\,arbitrary.
\item[(b)]It tells nothing about the ordering of neutrino masses, nor about its spectrum\,(degenerate or non-degenerate). If it says something in this regard, then this implies the framework is model dependent.
\item[(c)] The $|\nu_{3}\rangle=(0, 1/\sqrt{2},1/\sqrt{2})^T $ is the important signature of $2$-$3$ symmetry.
\item[(d)] A nonzero $\theta_{13}$ can be produced only if $2$-$3$ symmetry is broken.
\item[(e)] The mass eigenvalues and CP phases hardly interfere in defining the $2$-$3$ symmetric texture.
\end{itemize}    
The present study will reveal that these properties of $2$-$3$ symmetry are just a part of the whole picture and not the full portrait.

The $2$-$3$ symmetry discussed in this article is often termed as ``$\mu$-$\tau$ permutation symmetry'' and the present study is solely dedicated to investigate the same. There are also two different versions of $2$-$3$ symmetries: one is``$\mu$-$\tau$ antisymmetry''\cite{Kitabayashi:2005fc,Grimus:2005jk} and ``$\mu$-$\tau$ reflection symmetry''\cite{Harrison:2002et,Zhou:2014sya} which in general can shelter a nonzero $\theta_{13}$. 

\section{Searching other properties of $2$-$3$ symmetry}

Instead of getting into the confusion that may arise because of the parametrization adopted in Eqs\,(\ref{U1}) and (\ref{U2}), we start from the basic definition of $2$-$3$ symmetric texture of both kinds. We have,
\begin{equation}
\begin{pmatrix}
A & B & B\\
B & C & D\\
B & D & C
\end{pmatrix}_{\text{1st}}\,\,\text{and}\,\,\begin{pmatrix}
A & B & -B\\
B & C & D\\
-B & D & C
\end{pmatrix}_{\text{2nd}}.
\end{equation} 
The texture of the first kind is related to the invariance of the neutrino mass matrix under the transformation $|\nu_{\mu}\rangle \rightarrow |\nu_{\tau}\rangle $ and the second one with $|\nu_{\mu}\rangle \rightarrow -|\nu_{\tau}\rangle $\,\cite{Lam:2001fb}.
For definiteness, we shall follow the PMNS matrix, $U$ as such,
\begin{equation}
\label{U}
U=\left(
\begin{array}{ccc}
 c_{12} c_{13} \pmb{\text{}} & c_{13} s_{12} \pmb{\text{}} & e^{-i \delta } s_{13} \pmb{\text{}} \\
 -\frac{s_{12} \pmb{\text{}}}{\sqrt{2}}-\frac{e^{i \delta } c_{12} s_{13} \pmb{\text{}}}{\sqrt{2}} & \frac{c_{12} \pmb{\text{}}}{\sqrt{2}}-\frac{e^{i \delta } s_{12} s_{13} \pmb{\text{}}}{\sqrt{2}} & \frac{c_{13} \pmb{\text{}}}{\sqrt{2}} \\
 \frac{s_{12} \pmb{\text{}}}{\sqrt{2}}-\frac{e^{i \delta } c_{12} s_{13} \pmb{\text{}}}{\sqrt{2}} & -\frac{c_{12} \pmb{\text{}}}{\sqrt{2}}-\frac{e^{i \delta } s_{12} s_{13} \pmb{\text{}}}{\sqrt{2}} & \frac{c_{13} \pmb{\text{}}}{\sqrt{2}} \\
\end{array}
\right).P
\end{equation}
where rotations are taken as anti-clockwise. The $\theta_{23}$ is assumed as $45^0$ which is consistent within $2\sigma$ bound of global data. With the above $U$, we invert the equation\,(\ref{mnu}), to get the neutrino mass matrix as,
\begin{equation}
m_{\nu}=U.\,diag\lbrace m_{1},m_{2},m_{3}\rbrace U^{T}.
\end{equation}
We present $m_{\nu}$ in terms of a $2$-$3$ symmetric texture of first  kind followed by a deviation matrix as in the following,
\begin{equation}
\label{m1}
m_{\nu}= \left(
\begin{array}{ccc}
 \mathcal{A} & \mathcal{B} & \mathcal{B}\\
 \mathcal{B} & \mathcal{C} & \mathcal{D}\\
 \mathcal{B}& \mathcal{D} & \mathcal{C}
\end{array}\right)+ \left(
\begin{array}{ccc}
 0 & 0 & -\Delta_{1}\\
 0 & 0 & 0\\
 -\Delta_{1}& 0 & -\Delta_{2}
\end{array}
\right),
\end{equation}     
where,
\begin{eqnarray}
\label{d1}
\Delta_{1} &=& \frac{1}{\sqrt{2}}\sin 2 \theta _{12} \cos \theta _{13} \left(e^{2 i \beta } m_2-e^{2 i \alpha } m_1\right) ,\\
\label{d2}
\Delta_{2} &=& e^{i \delta } \sin 2 \theta _{12} \sin \theta _{13} \left(e^{2 i \alpha } m_1-e^{2 i \beta } m_2\right).
\end{eqnarray}
Similarly in terms of deviation from second kind $2$-$3$ symmetric texture we have,
\begin{equation}
\label{m2}
m_{\nu}= \left(
\begin{array}{ccc}
\mathcal{A}' & \mathcal{B}' & -\mathcal{B}'\\
 \mathcal{B}' & \mathcal{C}' & \mathcal{D}'\\
 -\mathcal{B}' & \mathcal{D}' & \mathcal{C}'
\end{array}
\right)+\left(
\begin{array}{ccc}
 0 & 0 & \Delta_{1}'\\
 0 & 0 & 0\\
 \Delta_{1}'& 0 & -\Delta_{2}'
\end{array}
\right);
\end{equation}
with deviation factors shown below, 
\begin{eqnarray}
\label{d1'}
\Delta_{1}' &=&-\frac{1}{\sqrt{2}} e^{-i \delta } \sin 2\theta _{13} (m_1 e^{2 i (\alpha +\delta )} \cos ^2\theta _{12}\nonumber\\
&&+m_2 e^{2 i (\beta +\delta )} \sin ^2\theta _{12}-m_3),\\
\label{d2'}
\Delta_{2}' &=& e^{i \delta } \sin 2 \theta _{12} \sin \theta _{13} \left(e^{2 i \alpha } m_1-e^{2 i \beta } m_2\right).
\end{eqnarray}
The details of the mass elements are presented in Tables\,(\ref{t1})-(\ref{t5}). Once the $\Delta$\,s and $\Delta'$\,s disappear, the exact $2$-$3$ symmetry is restored. These deviation factors may disappear in several ways ( see Tables\,(\ref{t2})-(\ref{t6})) and each route associate $2$-$3$ symmetry with newer aspects.

Note that if $\theta_{13}=0$, both $\Delta_1'$ and $\Delta_2'$ vanish (see Eqs\,(\ref{d1'}) and (\ref{d2'})), and second kind $2$-$3$ symmetric texture is achieved by $m_{\nu}$. The corresponding diagonalizing matrix, $U$ in Eq\,(\ref{U}) converges to the texture as shown in Eq\,(\ref{U2}). This treatment is consistent with the relation of second kind $2$-$3$ symmetric neutrino mass matrix and the pattern of $U$ as per the parametrization in Eq\,(\ref{U2}). What we believe generally is that the $2$-$3$ symmetric texture of first kind can hardly be related to $U$ of Eq\,(\ref{U2}). But the Eqs\,(\ref{d1}) and (\ref{d2}) show that this is in fact possible, where $\Delta_{1}$ varies as $\sin\theta_{13}$, and hence on setting $\theta_{13}=0$, one can get $\Delta_1=0$. Since $\Delta_{2}$ varies as $\cos\theta_{13}$ we require additional condition to make $\Delta_2=0$. These conditions may include the possibilities as such: $\sin2\theta_{12}=0$ or $m_1 e^{2i\alpha}=m_2 e^{2i\beta}$.

\subsection{Achieving $2$-$3$ symmetric texture of first kind without $\theta_{13}\neq 0$.}

We shall exemplify a few cases. One sees that in Eqs\,(\ref{d1}) and (\ref{d2}), even if $\theta_{13}\neq 0$, the condition, 
\begin{equation}
\label{eq1}
e^{2 i \beta } m_1-e^{2 i \alpha } m_2=0,
\end{equation}
is sufficient to bring about $2$-$3$ symmetry of first kind . This however implies,
\begin{equation}
\beta = \alpha-\frac{1}{2}\,i\,\log\left( \frac{m_1}{m_2}\right).
\end{equation}
Hence, only if $m_{1}=m_{2}$, we get $\beta=\alpha$ and vice versa . This $2$-$3$ symmetric platform forbids the necessity of zero reactor angle. However, it demands an exact degeneracy of the first two mass eigenstates, $|\nu_{1}\rangle$ and $|\nu_{2}\rangle$. We observe that out of nine parameters, the five parameters $(\theta_{12},\,\theta_{13},\,\alpha,\,\delta,\,m_1,m_3)$ are free. The neutrino mass matrix satisfying these conditions is presented below,
\begin{equation}
m_{\nu}=\begin{pmatrix}
A & B & B\\
B & C & D\\
B & D & C 
\end{pmatrix}
\end{equation}
where,

\begin{eqnarray}
A &=&  e^{2 i \alpha } m_1 \cos ^2\theta _{13}+e^{-2 i \delta } \sin ^2\theta _{13} m_3 ,\\
B &=& -\frac{e^{-i \delta } \sin 2 \theta _{13} \left(e^{2 i (\alpha +\delta )} m_1-m_3\right)}{2 \sqrt{2}} ,\\
C &=& \frac{1}{2} \left(m_3 \cos ^2\theta _{13}+e^{2 i \alpha } \left(e^{2 i \delta } \sin ^2\theta _{13}+1\right) m_1\right) ,\\
D &=& \frac{1}{2} \left(m_3 \cos ^2\theta _{13}+e^{2 i \alpha } \left(e^{2 i \delta } \sin ^2\theta _{13}-1\right) m_1\right).
\end{eqnarray}

Another exact $2$-$3$ symmetric texture is realized if $\sin2\theta_{12}=0$, i.e $\theta_{12}$ is considered as $90^0$. This is also an independent condition that leads to vanishing $\Delta_1$ and $\Delta_2$, and hardly requires $\theta_{13}=0$. This platform keeps all the other physical parameters (except $\theta_{23}$) free.  

Also one can think of the conditions like: $\theta_{13}=\pi/2$ and $\theta_{12}=\pi/2$\,(or, $e^{2 i \beta } m_1=e^{2 i \alpha } m_2$) to get the concerned symmetry.

\subsection{The $2$-$3$ symmetric texture of second kind.}

As stated earlier, if $\theta_{13}$ is zero, the $\Delta'$\,s in Eqs\,(\ref{d1'}) and (\ref{d2'}) become zero and all the parameters including $\theta_{12}$ are arbitrary. Hence, this scenario coincides with what the general perspective towards $2$-$3$ symmetry demands. 

We see the realization of $2$-$3$ symmetry if one associates Eqs\,(\ref{d1'}) and (\ref{d2'}) with the following constraints, 
\begin{equation}
\quad \sin2\theta_{13}=0,\quad m_{1}e^{2i\alpha}=m_{2}e^{2i\beta}.
\end{equation} 
Hence, this picture needs $\theta_{13}$ as $\pi/2$ for $2$-$3$ symmetry. The second condition hints for the exact degeneracy of $m_1$ and $m_2$ (and $\beta=\alpha$).

Let us look into other possibilities. If we say,
\begin{equation}
\label{c1}
\left(m_1 e^{2 i (\alpha +\delta )} \cos ^2\theta _{12}+m_2 e^{2 i (\beta +\delta )} \sin ^2\theta _{12}-m_3\right)=0,
\end{equation}
then the above condition will make  $\Delta_{1}'=0$. But in order to make, $\Delta_{2}'=0$, one of the following three possibilities is required, 
\begin{itemize}
\item[(a)]$m_{1}e^{2i\alpha}=m_{2}e^{2i\beta}$,  
\item[(b)]$\sin 2\theta_{12}=0$, and
\item[(c)]$\sin\theta_{13}=0$.
\end{itemize}

If we proceed with the condition(a), one sees,
\begin{eqnarray}
\alpha &=& -\delta-i\frac{1}{2}\log\left( \frac{m_3}{m_1}\right),\\
\beta &=& -\delta-i\frac{1}{2}\log\left( \frac{m_3}{m_2}\right).
\end{eqnarray}
One can see that if $m_1=m_2=m_3$, then $\alpha=\beta=-\delta$ and vice versa. This $2$-$3$ symmetric platform describes a highly degenerate spectrum of neutrino masses. Also, the CP phases are discerned to have the same magnitude. In this picture, out of five parameters only four: ($m-1,\,\delta,\,\theta_{13},\,\theta_{12}$) are free.

Note that the condition (c) $\sin\theta_{13}=0$  does not constraint the Eq\,(\ref{c1}) by no means, but on dissociating the real and imaginary parts of the latter, one arrives at the following relations,   
\begin{equation}
\label{f1}
\tan^2\theta_{12}=-\left(\frac{m_1}{m_2}\right)\mathcal{T},
\end{equation}
and,
\begin{equation}
\label{f4}
m_{3}=\frac{m_1 m_2 (\cos (2 \alpha +2 \delta )-\sin (2 \alpha +2 \delta ) \cot (2 \beta +2 \delta ))}{m_2-m_1 \sin (2 \alpha +2 \delta ) \csc (2 \beta +2 \delta )}.
\end{equation}
where,
\begin{equation}
\label{f2}
\mathcal{T}=\frac{\sin 2\left(\alpha + \delta \right)}{\sin 2\left(\beta+\delta \right)}
\end{equation}
The positivity of $\tan^2\theta_{12}$ demands, $\mathcal{T}<0$. We know that the Gatto Sartori Toninn (GST) relation is an important observation\,\cite{Gatto:1968ss} that relates the $1$-$2$ rotation angle of quark mixing matrix to the ratio of down type and strange quark masses.
\begin{equation}
\tan\theta_{C}\approx\theta_{C}=\sqrt{\frac{m_d}{m_s}}.
\end{equation} 
We see that Eq\,(\ref{f1}) is interesting in the sense, it gives a similar flavor of a GST relation even in the lepton sector\cite{Fritzsch:2006sm}.

\section{The $2$-$3$ symmetry of $m_{\nu}m_{\nu}^{\dagger}$}

Interesting would be the case if one tries to relate the different possibilities associated with $2$-$3$ symmetry in the context of $h=m_{\nu}m_{\nu}^{\dagger}$. The matrix, $h$ does not involve the information of two Majorana phases. On inverting Eq\,(\ref{h}) we construct $h$ and represent it in terms of an expression deviated from $2$-$3$ symmetric textures of either kind.

We see both cases,
\begin{eqnarray}
\label{mh1}
h=\begin{pmatrix}
A_{h} & B_{h} & B_{h}\\
B_{h} & C_{h} & D_{h}\\
B_{h} & D_{h} & C_{h}
\end{pmatrix}+\begin{pmatrix}
0 & 0 & -\Delta_{1h}\\
0 & 0 & 0\\
-\Delta_{1h} & 0 &-\Delta_{2h}
\end{pmatrix}.
\end{eqnarray}
where,
\begin{eqnarray}
\Delta_{1h} &=& \frac{1}{\sqrt{2}}\left(m_2^2-m_1^2\right) \sin 2 \theta _{12}\cos \theta _{13},\\
\Delta_{2h} &=& \left(m_1^2-m_2^2\right) \cos \delta  \sin 2 \theta _{12} \sin \theta _{13} .
\end{eqnarray}
and,
\begin{eqnarray}
\label{mh2}
h=\begin{pmatrix}
A_{h}' & B_{h}' & -B_{h}'\\
B_{h}' & C_{h}' & D_{h}'\\
-B_{h}' & D_{h}' & C_{h}'
\end{pmatrix}+\begin{pmatrix}
0 & 0 & \Delta'_{1h}\\
0 & 0 & 0\\
\Delta'_{1h} & 0 &-\Delta'_{2h}
\end{pmatrix}.
\end{eqnarray}
where,
\begin{eqnarray}
\Delta_{1h}' &=& \frac{1}{\sqrt{2}} e^{-i \delta } \sin 2\theta _{13} \left(m_3^2-m_2^2 \sin ^2\theta _{12}-m_1^2 \cos ^2\theta _{12}\right),\nonumber \\
\Delta_{2h}' &=& \left(m_1^2-m_2^2\right) \cos \delta \sin 2 \theta _{12} \sin \theta _{13}.
\end{eqnarray}

The description of the concerned matrix elements are tabulated in Tables\,(\ref{t3}) and (\ref{t7}). If $m_{\nu}$ is $2$-$3$ symmetric then $h$ has to follow the same symmetry. One can see that the first kind $2$-$3$ symmetric texture of $h$ is achievable if $m_{2}=m_{1}$ and the second kind demands $m_1=m_2=m_3$. The requirements are consistent with whatever is discussed in earlier section. The various possibilities in order to obtain the $2$-$3$ symmetric textures of $h$ are depicted in Tables\,(\ref{t4}) and (\ref{t8}).

\begin{table*}
\begin{center}
\setlength{\tabcolsep}{0.2 em}
\begin{tabular}{c l}
\hline 
\hline
$\mathcal{A}=$&$\cos ^2\theta _{13} \left(e^{2 i \alpha } m_1 \cos ^2\theta _{12}+e^{2 i \beta } m_2 \sin ^2\theta _{12}\right)+e^{-2 i \delta } m_3 \sin ^2\theta _{13}$\\
$\mathcal{B}=$ &\begin{tabular}{l}
$-\frac{1}{\sqrt{2}}e^{-i \delta }\cos \theta _{13}\lbrace m_1 e^{i (2 \alpha +\delta )} \cos \theta _{12} \left(\sin \theta _{12}+e^{i \delta } \sin \theta _{13} \cos \theta _{12}\right)+ m_2 e^{i (2 \beta +\delta )} \sin \theta _{12} \left(e^{i \delta } \sin\theta _{12} \sin \theta _{13}-\cos \theta _{12}\right)-m_3 \sin \theta _{13}\rbrace$
\end{tabular} \\
$\mathcal{C}=$ &\begin{tabular}{l}
$\frac{1}{2}\lbrace e^{2 i \alpha } m_1 \left(\sin \theta _{12}+e^{i \delta } \sin \theta _{13} \cos \theta _{12}\right){}^2+ m_3 \cos ^2\theta _{13}+e^{2 i \beta } m_2 \left(\cos \theta _{12}-e^{i \delta } \sin \theta _{12} \sin \theta _{13}\right){}^2\rbrace$
\end{tabular} \\
$\mathcal{D}=$ &\begin{tabular}{l}
$\frac{1}{2}\lbrace -e^{2 i \alpha } m_1 \left(\sin ^2\theta _{12}-e^{2 i \delta } \sin ^2\theta _{13} \cos ^2\theta _{12}\right)-e^{2 i \beta } m_2 \left(\cos ^2\theta _{12}-e^{2 i \delta } \sin ^2\theta _{12} \sin ^2\theta _{13}\right)m_3 +\cos ^2\theta _{13}\rbrace$
\end{tabular} \\
\hline 
\end{tabular}
\caption{\label{t1}\footnotesize The description of the matrix elements in the $2$-$3$ symmetric texture of first kind in Eq\,(\ref{m1}) .} 
\end{center}
\end{table*}
\begin{table*}
\begin{center}
\setlength{\tabcolsep}{0.2 em}
\begin{tabular}{cl}
\hline 
\hline
$\mathcal{A}'=$&$\cos ^2\theta _{13} \left(e^{2 i \alpha } m_1 \cos ^2\theta _{12}+e^{2 i \beta } m_2 \sin ^2\theta _{12}\right)+e^{-2 i \delta } m_3 \sin ^2\theta _{13}$\\
$\mathcal{B}'=$ &\begin{tabular}{l}
$-\frac{1}{\sqrt{2}}e^{-i \delta }\cos \theta _{13}\lbrace m_1 e^{i (2 \alpha +\delta )} \cos \theta _{12} \left(\sin \theta _{12}+e^{i \delta } \sin \theta _{13} \cos \theta _{12}\right)-m_3 \sin \theta _{13}+m_2 e^{i (2 \beta +\delta )} \sin\theta _{12} \left(-\cos\theta _{12}+e^{i \delta } \sin \theta _{12} \sin \theta _{13}\right)\rbrace$
\end{tabular} \\
$\mathcal{C}'=$ &\begin{tabular}{l}
$\frac{1}{2}\lbrace e^{2 i \alpha } m_1 \left(\sin \theta _{12}+e^{i \delta } \sin \theta _{13} \cos \theta _{12}\right){}^2 m_3 \cos ^2\theta _{13}+e^{2 i \beta } m_2 \left(\cos \theta _{12}-e^{i \delta } \sin \theta _{12} \sin \theta _{13}\right){}^2\rbrace$
\end{tabular} \\
$\mathcal{D}'=$ &\begin{tabular}{l}
$\frac{1}{2}\lbrace -e^{2 i \alpha } m_1 \left(\sin ^2\theta _{12}-e^{2 i \delta } \sin ^2\theta _{13} \cos ^2\theta _{12}\right)+m_3 \cos ^2\theta _{13}-e^{2 i \beta } m_2 \left(\cos ^2\theta _{12}-e^{2 i \delta } \sin ^2\theta _{12} \sin ^2\theta _{13}\right)\rbrace$
\end{tabular} \\
\hline 
\end{tabular}
\caption{\label{t5}\footnotesize The description of the matrix elements in the $2$-$3$ symmetric texture of second kind appearing in Eq\,(\ref{m2}) .} 
\end{center}
\end{table*}
\begin{table*}
\begin{center}
\setlength{\tabcolsep}{2 em}
\begin{tabular}{c|c c}
\hline
\hline 
$m_{\nu}$ & Conditions Required & Free parameters involved \\ 
\hline
\hline
$\left(
\begin{array}{lll}
 \mathcal{A} & \mathcal{B} & \mathcal{B}\\
 \mathcal{B} & \mathcal{C} & \mathcal{D}\\
 \mathcal{B}& \mathcal{D} & \mathcal{C}
\end{array}\right)$ & \begin{tabular}{l}
$\sin\theta_{13}=0,\sin 2\theta_{12}=0$ \\ 
$\sin\theta_{13}=0,\left(e^{2 i \beta } m_1-e^{2 i \alpha } m_2\right) =0$\\
$\sin 2\theta_{12} = 0$\\
$\left(e^{2 i \beta } m_1-e^{2 i \alpha } m_2\right) = 0$\\
$\cos\theta_{13}=0,\sin 2\theta_{12}=0$ \\
$\cos\theta_{13}=0, \left(e^{2 i \beta } m_1-e^{2 i \alpha } m_2\right) =0$\\
\end{tabular}  & \begin{tabular}{l}
$m_{1},m_{2},m_{3},\delta,\alpha,\beta$\\

$m_{1},m_{3},\theta_{12},\delta,\alpha$\\
 
$m_{1},m_{2},m_{3},\theta_{13},\alpha,\beta,\delta$\\
 
$m_{1},m_{3},\theta_{13},\delta,\alpha$\\
 
$m_{1},m_2,m_{3},\delta,\alpha,\beta$\\
 
$m_{1},m_{3},\theta_{12},\delta,\alpha$
\end{tabular}\\
\hline 
\end{tabular} 
\caption{\label{t2}\footnotesize The different conditions that can give rise to $2$-$3$ symmetric texture of first kind.} 
\end{center}
\end{table*}
\begin{table*}
\begin{center}
\setlength{\tabcolsep}{0.15 em}
\begin{tabular}{c|c l}
\hline
\hline 
$m_{\nu}$ & Conditions Required & Free parameters \\ 
\hline
\hline
$\left(
\begin{array}{ccc}
 \mathcal{A}' & \mathcal{B}' & -\mathcal{B}'\\
 \mathcal{B}' & \mathcal{C}' & \mathcal{D}'\\
 -\mathcal{B}' & \mathcal{D}' & \mathcal{C}'
\end{array}\right)$ & \begin{tabular}{l}
$\sin\theta_{13}=0$ \\ 
$\sin 2\theta_{12}=0,(m_1 e^{2 i (\alpha +\delta )} \cos ^2\theta _{12}+m_2 e^{2 i (\beta +\delta )} \sin ^2\theta _{12}-m_3)=0$\\
$\sin2\theta_{13}=0,(e^{2 i \alpha } m_1-e^{2 i \beta } m_2)=0$\\
$(m_1 e^{2 i (\alpha +\delta )} \cos ^2\theta _{12}+m_2 e^{2 i (\beta +\delta )} \sin ^2\theta _{12}-m_3)=0,(e^{2 i \alpha } m_1-e^{2 i \beta } m_2)=0$\\
$\sin\theta_{13}=0,(m_1 e^{2 i (\alpha +\delta )} \cos ^2\theta _{12}+m_2 e^{2 i (\beta +\delta )} \sin ^2\theta _{12}-m_3)=0$ \\
\end{tabular}  & \begin{tabular}{l}
$m_{1},m_{2},m_{3},\theta_{12},\delta,\alpha,\beta$\\
$m_{1},m_{2},\theta_{13},\delta,\alpha$\\
$m_{1},m_{3},\alpha,\beta,\delta$\\
$m_{1},\theta_{13},\theta_{12},\delta$\\
$m_{1},m_2,\delta,\alpha,\beta$\\ 
\end{tabular} \\ 
\hline 
\end{tabular} 
\caption{\label{t6}\footnotesize The conditions that can lead to $2$-$3$ symmetric texture of second kind.} 
\end{center}
\end{table*}  

\begin{table*}
\begin{center}
\setlength{\tabcolsep}{1 em}
\begin{tabular}{c|l}
\hline 
\hline
$\mathcal{A}_h$ & $m_3^2 \sin ^2\theta _{13}+\cos ^2\theta _{13} \left(m_2^2 \sin ^2\theta _{12}+m_1^2 \cos ^2\theta _{12}\right) $\\
\\
$\mathcal{B}_{h}$ &\begin{tabular}{l}
 $-\frac{m_1^2 \sin 2 \theta _{12} \cos \theta _{13}}{2 \sqrt{2}}-\frac{e^{-i \delta } m_1^2 \sin 2\theta _{13} \cos ^2\theta _{12}}{2\sqrt{2}} +\frac{m_2^2 \sin 2 \theta _{12} \cos \theta _{13}}{2 \sqrt{2}}$\\ 
 $+\frac{e^{-i \delta } m_3^2 \sin 2\theta _{13}}{2\sqrt{2}}-\frac{e^{-i \delta } m_2^2 \sin ^2 \theta _{12} \sin 2\theta _{13}}{2\sqrt{2}}$
\end{tabular} \\
\\
$\mathcal{C}_h$ &\begin{tabular}{l}
$\frac{1}{8}\lbrace 4 m_1^2 \left(\sin \theta _{13} \left(\cos\delta \sin 2 \theta _{12}+\sin \theta _{13} \cos ^2\theta _{12}\right)+\sin ^2\theta _{12}\right)$\\ 
$+m_2^2 \left(-4 \cos \delta  \sin 2 \theta _{12} \sin \theta _{13}+\cos 2 \theta _{12}-2 \sin ^2\theta _{12} \cos 2 \theta _{13}+3\right)+4 m_3^2 \cos ^2\theta _{13}\rbrace $
\end{tabular} \\
\\
$\mathcal{D}_h$ &\begin{tabular}{l}
$\frac{1}{4}e^{-i \delta }\lbrace m_1^2 \left(-2 e^{i \delta } \sin ^2\theta _{12}-\left(e^{2 i \delta }-1\right) \sin 2 \theta _{12} \sin \theta _{13}+2 e^{i \delta } \sin ^2\theta _{13} \cos ^2\theta _{12}\right)$\\ 
 $+ 2 e^{i \delta } m_3^2 \cos ^2\theta _{13}+m_2^2 \left(2 e^{i \delta } \sin ^2\theta _{12} \sin ^2\theta _{13}+\left(e^{2 i \delta }-1\right) \sin 2 \theta _{12} \sin\theta _{13}-2 e^{i \delta } \cos ^2\theta _{12}\right)\rbrace$
\end{tabular} \\
\hline
\end{tabular}
\caption{\label{t3}\footnotesize The description of the matrix elements of first kind $2$-$3$ symmetric matrix appearing in the texture $h=m_{\nu}m_{\nu}^{\dagger}$ in the Eq.\,(\ref{mh1}).} 
\end{center}
\end{table*}

\begin{table*}
\begin{center}
\setlength{\tabcolsep}{4 em}
\begin{tabular}{c|c}
\hline
\hline 
$h=m_{\nu}m_{\nu}^{\dagger}$ & Conditions Required \\ 
\hline
\hline
$\left(
\begin{array}{ccc}
 \mathcal{A}_{h} & \mathcal{B}_{h} & \mathcal{B}_{h}\\
 \mathcal{B}_{h} & \mathcal{C}_{h} & \mathcal{D}_{h}\\
 \mathcal{B}_{h}& \mathcal{D}_{h} & \mathcal{C}_{h}
\end{array}\right)$ & \begin{tabular}{l}
$\sin\theta_{13}=0,\sin2\theta_{12}=0$ \\ 
$\sin\theta_{13}=0,\left(m_{2}^2-m_1^2 \right)=0$\\
$\cos\delta=0, \cos\theta_{13}=0$\\
$\cos\delta = 0, \left(m_{2}^2-m_1^2 \right)=0$\\
$\sin2\theta_{12}=0$ \\
$\left(m_{2}^2-m_1^2 \right)=0$
\end{tabular} \\
\hline
\end{tabular} 
\caption{\label{t4}\footnotesize The conditions to obtain a first kind $2$-$3$ symmetric $h$. .} 
\end{center}
\end{table*}

\begin{table*}
\begin{center}
\setlength{\tabcolsep}{0.2 em}
\begin{tabular}{c|l}
\hline 
\hline
$\mathcal{A}_h'$ & $m_3^2 \sin ^2\theta _{13}+\cos ^2\theta _{13} \left(m_2^2 \sin ^2\theta _{12}+m_1^2 \cos ^2\theta _{12}\right) $\\
$\mathcal{B}_{h}'$ &\begin{tabular}{l}
 $ \frac{1}{2 \sqrt{2}}\lbrace e^{-i \delta } \cos \theta _{13} \left(-2 \sin \theta _{13} \left(m_2^2 \sin ^2\theta _{12}+m_1^2 \cos ^2\theta _{12}-m_3^2\right)-e^{i \delta } \left(m_1^2-m_2^2\right) \sin 2 \theta _{12}\right)\rbrace$\\ 
\end{tabular} \\
$\mathcal{C}_h'$ &\begin{tabular}{l}
{\footnotesize$ \frac{1}{8} \lbrace 4 m_1^2 \left(\sin \theta _{13} \left(\cos \delta  \sin 2 \theta _{12}+\sin\theta _{13} \cos ^2\theta _{12}\right)+\sin ^2\theta _{12}\right)$}\\ 
{\footnotesize$ +m_2^2 \left(-4 \cos \delta \sin 2 \theta _{12} \sin \theta _{13}+\cos 2 \theta _{12}-2 \sin ^2\theta _{12} \cos 2 \theta _{13}+3\right)+4 m_3^2 \cos ^2\theta _{13}\rbrace$}
\end{tabular} \\
$\mathcal{D}_h'$ &\begin{tabular}{l}
{\footnotesize$ \frac{1}{4} e^{-i \delta } \lbrace (2 e^{i \delta } m_3^2 \cos ^2\left(\theta _{13}\right)+m_1^2 \left(-2 e^{i \delta } \sin ^2\left(\theta _{12}\right)-\left(-1+e^{2 i \delta }\right) \sin \left(2 \theta _{12}\right) \sin \left(\theta _{13}\right)+2 e^{i \delta } \sin ^2\left(\theta _{13}\right) \cos ^2\left(\theta _{12}\right)\right) $}\\ 
{\footnotesize $ +m_2^2 \left(2 e^{i \delta } \sin ^2\left(\theta _{12}\right) \sin ^2\left(\theta _{13}\right)+\left(-1+e^{2 i \delta }\right) \sin \left(2 \theta _{12}\right) \sin \left(\theta _{13}\right)-2 e^{i \delta } \cos ^2\left(\theta _{12}\right)\right)\rbrace$}
\end{tabular} \\
\hline 
\end{tabular}
\caption{\label{t7}\footnotesize The description of the matrix elements of $2$-$3$ symmetric matrix of second kind appearing in the texture of $h=m_{\nu}m_{\nu}^{\dagger}$ in the Eq.\,(\ref{mh2}).} 
\end{center}
\end{table*}

\begin{table*}
\begin{center}
\setlength{\tabcolsep}{1 em}
\begin{tabular}{c|c}
\hline
\hline 
$m_{\nu}m_{\nu}^{\dagger}$ & Conditions Required \\ 
\hline
\hline
$\left(
\begin{array}{ccc}
 \mathcal{A}_{h}' & \mathcal{B}_{h}' & -\mathcal{B}_{h}'\\
 \mathcal{B}_{h}' & \mathcal{C}_{h}' & \mathcal{D}_{h}'\\
 -\mathcal{B}_{h}'& \mathcal{D}_{h}' & \mathcal{C}_{h}'
\end{array}\right)$ & \begin{tabular}{l}
$\sin\theta_{13}=0 $ \\ 
$ \sin2\theta_{13}=0,\cos\delta=0$\\
$\left(m_2^2-m_1^2 \right)=0,\sin2\theta_{13}=0 $\\
$\sin2\theta_{13}=0,\sin 2\theta_{12}=0 $\\
$\left(m_2^2-m_1^2\right)=0,\left(m_3^2-m_2^2 \sin ^2\theta _{12}-m_1^2 \cos ^2\theta _{12}\right)=0 $ \\
$\cos\delta=0, \quad\left(m_3^2-m_2^2 \sin ^2\theta _{12}-m_1^2 \cos ^2\theta _{12}\right)=0 $\\
$\sin 2\theta_{12}=0, \quad\left(m_3^2-m_2^2 \sin ^2\theta _{12}-m_1^2 \cos ^2\theta _{12}\right)=0$
\end{tabular} \\
\hline
\end{tabular} 
\caption{\label{t8}\footnotesize The conditions to obtain a second kind $2$-$3$ symmetric $h$.} 
\end{center}
\end{table*}

\section{The $2$-$3$ symmetry and $\theta_{23}= 45^0$}

The discussion concerned with $2$-$3$ symmetry is based on the form of $U$ shown in Eq.\,(\ref{U}), where $\theta_{23}$ is fixed at $45^0$. As it is seen in our analysis that a non-vanishing $\theta_{13}$ can be related to the $2$-$3$ symmetry under certain occasion, one can question the popular choice of $\theta_{23}=45^0$ that seems to have an intimate relation with the same symmetry. For definiteness, we keep $\theta_{23}$ arbitrary and define $U$ upto the Dirac and Majorana phases as in the following,
\begin{equation}
U=\left(
\begin{array}{ccc}
 c_{13} c_{12} & c_{13} s_{12} & s_{13} \\
 -c_{12} s_{23} s_{13}-c_{23} s_{12} & c_{23} c_{12}-s_{23} s_{13} s_{12} & c_{13} s_{23} \\
 s_{23} s_{12}-c_{23} c_{12} s_{13} & -c_{12} s_{23}-c_{23} s_{13} s_{12} & c_{23} c_{13} \\
\end{array}
\right).
\end{equation}   
The neutrino mass matrix can be expressed as such,
\begin{eqnarray}
M_{\nu}=U.M_{\nu}^{d}.U^T=\begin{bmatrix}
A & B & -B\\
B & C & D\\
-B & D & C
\end{bmatrix}+\begin{bmatrix}
0 & 0 & \Delta_{1}\\
0 & 0 & 0\\
\Delta_{1} & 0 & \Delta_{2}
\end{bmatrix}
\end{eqnarray},
where,
\begin{small}
\begin{eqnarray}
\Delta_{1}&=& c_{13} (-c_{23} (c_{12}^2 m_1 s_{13}+c_{12} (m_1-m_2) s_{12}\nonumber\\
&& +s_{13}(m_2 s_{12}^2-m_3))-s_{23} (c_{12}^2 m_1 s_{13}+c_{12} (m_2-m_1) s_{12}\nonumber\\
& &+s_{13}(m_2 s_{12}^2-m_3))),\\
\Delta_{2} &=& -m_3 c_{13}^2 s_{23}^2-m_1 (c_{12} s_{23} s_{13}+c_{23} s_{12}){}^2\nonumber\\
&& +m_2 (c_{12} s_{23}+c_{23} s_{13} s_{12}){}^2\nonumber\\
&& +m_1 (c_{23} c_{12} s_{13}-s_{23} s_{12}){}^2-m_2 (c_{23} c_{12}-s_{23} s_{13} s_{12}){}^2\nonumber\\
&& +c_{23}^2 c_{13}^2 m_3.
\end{eqnarray}
\end{small}

We consider the following cases to illustrate,
\begin{itemize}
\item[(a)] If $\theta_{13}=0$, and $m_2\neq m_1$ then,
\begin{eqnarray}
\Delta_{1} &=& (m_2-m_1)(c_{23}-s_{23})s_{12} c_{12},\\
\Delta_{2} &=& (1-2s_{23}^2)(m_3 -m_1 s_{12}^2-m_2c_{12}^2).
\end{eqnarray}
\item[(b)] If $m_{1}=m_{2}=m$ and $\theta_{13}\neq 0$ then,
\begin{eqnarray}
\Delta_{1} &=& (m_3-m)(c_{23}-s_{23})s_{13}c_{13},\nonumber\\
\Delta_{2} &=& (m_3-m)(1-2s_{23}^2)c_{13}^2.
\end{eqnarray}
\end{itemize}
One sees that neither $\theta_{13}=0$ nor $m_{1}=m_2$ is sufficient to make $\Delta$'s zero. The only possibility that fulfills this task is,
\begin{eqnarray}
1-2s_{23}^2=0 \quad\text{or}\quad c_{23}-s_{23}=0. 
\end{eqnarray} 
Hence, $\theta_{23}$ has to be $45^0$ if $2$-$3$ symmetry is demanded. By considering the other scenarios too one can conclude the same. In other words, $\theta_{23}=45^0$ is a part and parcel of $2$-$3$ symmetry.

\section{Discussion}
The investigation so far highlights many possibilities with $2$-$3$ symmetry and emphasizes that a vanishing $\theta_{13}$ with an arbitrary $\theta_{12}$ is a very special case of $2$-$3$ symmetry. Although a nonzero $\theta_{13}$ can be associated with a $2$-$3$ symmetric texture, yet the latter  
does not allow any deviation from $\theta_{23}=45^0$. Once we deviate from the condition of $\theta_{23}=45^0$, the $2$-$3$ symmetry is wrenched. We summarize some important features of present discussion.
\begin{itemize}
\item[(a)]With both $\theta_{13}=90^0$ and $\theta_{12}=90^{0}$, one can experience $2$-$3$ symmetry.
\item[(b)]Out of the nine physical parameters, not only the mixing angles, even mass eigenvalues and CP phases play an important role in determining the $2$-$3$ symmetric nature. 
\item[(c)] Certain partial degenerate condition or full degenerate condition of the three neutrino masses can be related to $2$-$3$ symmetric pattern.
\item[(d)]$\theta_{23}=45^{0}$ is a necessary condition for $2$-$3$ symmetry\,\cite{Zhou:2014sya}.
\item[(e)]$\theta_{13}=0$, $\theta_{12}$ being arbitrary is a special case. Both $\theta_{12}$ and $\theta_{13}$ can be made arbitrary in the $2$-$3$ symmetric background.
\item[(f)] We observe that not only $|\nu_{3}\rangle=(0,1/\sqrt{2},1/\sqrt{2})^T $, but also $|\nu_{3}\rangle=(s_{13} e^{-i\delta}, c_{13}/\sqrt{2},c_{13}/\sqrt{2})^T $ can be associated to $2$-$3$ symmetry of neutrino mass matrix.
\item[(g)] The $2$-$3$ symmetry is hierarchy-blind, but it can be associated with a degenerate spectrum: either partial or full. 
\end{itemize}   

The present study is model independent. But we hope that it will help the model builders to visualize the $2$-$3$ symmetry from a deeper perspective. We emphasize on the following points.

\begin{itemize}
\item[(a)] The present data says, $\theta_{13}\approx 9^0$. But we have seen that this can be accommodated within $2$-$3$ symmetry easily. 

\item[(b)] We have studied different cases which do not require vanishing $\theta_{13}$ to maintain $2$-$3$ symmetry. Out of those the cases which shows $\Delta\,m_{21}^2=0$ (which signifies that $m_2$ and $m_1$ are exactly degenerate), seems interesting. The experimental data shows that $\Delta\,m_{21}^2\approx 7.5 \times 10^{-5}\,eV^2$. Hence one can not rule out the fact that $m_{2}\gtrsim m_{1}$. In this light, a $2$-$3$ symmetric framework with $m_2=m_1$ appears as a good approximation to start with.

\item[(c)] Realization of Similar GST relation in the lepton sector is interesting from Grand Unified Theory (GUT) point of view. This is the result of sum rule,  $(m_{\nu}')_{12}=-(m_{\nu}')_{13}$ which is essential for $2$-$3$ symmetry.

\item[(d)]The $2$-$3$ symmetry is realizable in high energetic condition\cite{Luo:2014upa} and following the Renormalization group equations one may engender the observable effects.  

\item[(e)] The present approach addresses not only the Dirac CP phase $\delta$ but also the Majorana CP phases\,($\alpha$ and $\beta$). In this regard, one sees certain correlation between the concerned phases.

\item[(f)] We see that the $2$-$3$ symmetric pattern of $m_{\nu}m_{\nu}^{\dagger}$ can be associated with a condition, $\cos\delta=0$, which indicates a maximal CP violation.

\end{itemize}

Here we emphasize that the present discussion concerning the $2$-$3$ symmetry is devoted only to see whether the platform can serve as a convincing first approximation or not. We know that observed best-fit value of $\theta_{23}$ is $43^0$ for normal ordering of neutrino masses. The $2$-$3$ symmetry prediction: $\theta_{23}=45^0$ falls within $1\sigma$. In reality, neither $\theta_{13}$ is vanishing nor the two neutrino masses are exactly degenerate. Hence the original $2$-$3$ symmetric has to be perturbed. But the detailed discussion presented in this article has brought to light several possibilities regarding the ways one can perturb the original framework. The $2$-$3$ symmetry is very often criticized because it predicts a $\theta_{13}$ that requires a correction of the order of Cabibbo angle\,\cite{Cabibbo:1977nk} which is large. But the present approach entertains one possibility to perturb $2$-$3$ symmetry not in terms of $\theta_{13}$ but in terms of the exact degeneracy of $m_{1}$ and $m_2$ instead of perturbing $\theta_{13}$. In the former scenario, perturbation is required to lift $\theta_{13}$ from zero to certain finite value, whereas it appears to deviate the parameters from their exact degenerate value. The second case seems  interesting in the sense that there is a possibility to think of a $2$-$3$ symmetric platform that may hold a $\theta_{13}$ which is either equal to $\theta_{C}$ or $\mathcal{O}(\theta_C)$. The bi-Large mixing\,\cite{Boucenna:2012xb, Ding:2012wh, Roy:2014nua}, modulated Bi-maximal mixing\,\cite{Roy:2015cka} and Tri-bimaximal Cabibbo mixing\,\cite{King:2012vj} schemes which are interesting from Grand unified theory point of view, can now be related to $2$-$3$ symmetric framework. One may adopt the renormalization group equation effects\cite{Zhou:2014sya} in order to generate the observable mass square difference . 

Its worth mentioning that the present discussion finds some similarity with the ref.\,\cite{Adhikary:2013bma}, where the authors have introduced one cyclic symmetry (not $2$-$3$ symmetry) of the neutrino mass matrix. 

In summary, we have established the fact that within $2$-$3$ symmetric framework, there are several possibilities to realize a nonzero $\theta_{13}$. We have pointed out that the $2$-$3$ symmetric texture of neutrino mass matrix is achievable not only at the cost of special choice of $U$, but the neutrino mass eigenvalues and CP phases may also play a leading role in defining this symmetry of the neutrino mass matrix. The present investigation justifies the relevance of $2$-$3$ symmetry of $m_{\nu}$ as a first approximation.

\bibliography{bibs}

\end{document}